\documentclass[10pt,a4paper]{article}
\usepackage[numbers,sort&compress]{natbib}
\usepackage{authblk}
\usepackage{graphicx}
\usepackage[whole]{bxcjkjatype} 
\usepackage[top=30truemm,bottom=30truemm,left=25truemm,right=25truemm]{geometry}
\usepackage[colorlinks=true,linktoc=page,citecolor=red,linkcolor=blue]{hyperref}
\graphicspath{{./figs/}}
\usepackage{color,comment}
\usepackage{amsmath,amssymb,amsfonts}
\usepackage{bm,braket,array}
\usepackage[all]{xy}
\usepackage{amsthm,amscd}
\usepackage{mdframed}
\usepackage{slashed}
\usepackage{multirow}
\usepackage[format=hang,justification=raggedright]{caption}

\theoremstyle{definition}

\theoremstyle{remark}

\newcommand{\Z}{\mathbb{Z}}

\newcommand{\bk}{{\bm{k}}}
\newcommand{\br}{{\bm{r}}}

\newcommand{\wb}[1]{%
  \mathchoice
    {\mkern2mu\overline{\mkern-2mu #1 \mkern-2mu}\mkern2mu}
    {\mkern2mu\overline{\mkern-2mu #1 \mkern-2mu}\mkern2mu}
    {\mkern1.5mu\overline{\mkern-1.5mu #1 \mkern-1.5mu}\mkern1.5mu}
    {\mkern1mu\overline{\mkern-1mu #1 \mkern-1mu}\mkern1mu}
}
\newsavebox{\foobox}

\begin{document}
\title{Classification of topological insulators and superconductors with multiple order-two point group symmetries}
\author{Ken Shiozaki}
\affil{Center for Gravitational Physics and Quantum Information, Yukawa Institute for Theoretical Physics, Kyoto University, Kyoto 606-8502, Japan}
\date{\today}
\maketitle
\begin{abstract}
We present a method for computing the classification groups of topological insulators and superconductors in the presence of $\mathbb{Z}_2^{\times n}$ point group symmetries, for arbitrary natural numbers $n$. Each symmetry class is characterized by four possible additional symmetry types for each generator of $\mathbb{Z}_2^{\times n}$, together with bit values encoding whether pairs of generators commute or anticommute. We show that the classification is fully determined by the number of momentum- and real-space variables flipped by each generator, as well as the number of variables simultaneously flipped by any pair of generators. As a concrete illustration, we provide the complete classification table for the case of $\mathbb{Z}_2^{\times 2}$ point group symmetry.
\end{abstract}

\section{Introduction}

The classification of topological insulators and superconductors was initially understood in terms of internal symmetries such as time-reversal and particle–hole symmetries~\cite{SchnyderClassification2008,KitaevPeriodic2009,RyuTopological2010}. 
It was later extended by incorporating crystalline symmetries~\cite{TeoSurface2008,FuTopological2011,HsiehTopological2012}, leading to the establishment of the concept of higher-order topological insulators and superconductors, characterized by boundary gapless states localized not on surfaces or edges but at corners or hinges~\cite{BenalcazarQuantized2017,BenalcazarElectric2017,FangNew2019,SchindlerHigherorder2018,SchindlerHigherorder2018a,Song$densuremath2$Dimensional2017,LangbehnReflectionSymmetric2017}. These developments have greatly broadened the understanding of symmetry-protected topological phases.

In this direction, the $K$-theoretic framework developed by Freed and Moore~\cite{FreedTwisted2013,ThiangKTheoretic2016} has provided a general setting for the classification of topological insulators and superconductors with arbitrary crystalline symmetries. As a computational tool for $K$-theory, the Atiyah–Hirzebruch spectral sequence has been refined into a powerful and systematic method for deriving classifications~\cite{ShiozakiAtiyahHirzebruch2022,ShiozakiGeneralized2023,ShiozakiAtiyahHirzebruch2023}. On the other hand, for topological insulators and superconductors protected solely by (magnetic) point group symmetries without lattice translations, Cornfeld and Chapman introduced the on-site reduction of point group symmetries~\cite{CornfeldClassification2019}, which, together with subsequent developments~\cite{ShiozakiClassification2022}, established a concrete method to compute the classification groups.

In this paper, we pursue this comprehensive line of study and focus on the classification of topological insulators and superconductors with multiple $\mathbb{Z}_2$ point group symmetries. While the case of a single $\mathbb{Z}_2$ point group symmetry has already been understood~\cite{ShiozakiTopology2014}, a systematic understanding of the situation with several simultaneous $\mathbb{Z}_2$ symmetries has been lacking. Although general computational methods for classification groups with arbitrary point group symmetries have been established~\cite{CornfeldClassification2019,ShiozakiClassification2022}, our approach provides a way to determine the relevant $K$-groups explicitly in terms of a small number of parameters. This offers practical utility, for example, in systematic classification calculations of higher-order topological insulators and superconductors under $\mathbb{Z}_2$ point group symmetries~\cite{TrifunovicHigherOrder2019}. Specifically, by employing suspension isomorphisms in $K$-theory~\cite{TeoTopological2010}, we reduce the problem to the classification groups in zero dimension, thereby revealing the hierarchical structure of the classification with arbitrary numbers of $\mathbb{Z}_2$ point group symmetries. As a concrete example, we explicitly compute and present the classification table for the case of $\mathbb{Z}_2^{\times 2}$ point group symmetry.

\section{Formulation}
\subsection{Notation for symmetries and $K$-groups}

We first compute how the $K$-group in the presence of the real Altland-Zirnbauer (AZ) class and $n$ additional unitary $\mathbb{Z}_2$ symmetries can be reduced to the corresponding zero-dimensional $K$-group. As the full parameter space, we consider momentum-type variables $\bk = (k_1,\dots,k_d)$ and real-space variables surrounding a codimension $(D-1)$-defect $\br = (r_1,\dots,r_D)$, and compactify $\mathbb{R}^{d+D}$ into the $(d+D)$-dimensional sphere $S^{d+D}$. 

The eight real AZ classes are labeled by an integer $s \in \{0,\dots,7\}$, and we introduce the binary representation $s = (s_2s_1s_0)_2$, i.e., 
\begin{align}
s = 4s_2 + 2s_1 + s_0, \quad s_0, s_1, s_2 \in \{0,1\}.
\label{eq:s_2adic}
\end{align}
The Hamiltonian $H(\bk,\br)$ has the following symmetries depending on the AZ class:
\begin{align}
\text{$s_0=0$ (non-chiral class)}: \quad 
A H(\bk,\br) A^{-1} = (-1)^{s_1} H(-\bk,\br), \quad A^2 = (-1)^{s_2},
\label{eq:AZ_sym_nonchiral}
\end{align}
\begin{align}
\text{$s_0=1$ (chiral class)}: \quad 
\begin{cases}
A H(\bk,\br) A^{-1} = (-1)^{s_1} H(-\bk,\br), \quad A^2 = (-1)^{s_2}, \\
\Gamma H(\bk,\br) \Gamma^{-1} = - H(\bk,\br), \quad \Gamma^2 = 1, \\
A \Gamma = (-1)^{s_1} \Gamma A.
\end{cases}
\label{eq:AZ_sym_chiral}
\end{align}
Here $\Gamma$ and $A$ denote unitary and antiunitary operators, respectively.  

In addition, we consider unitary $\mathbb{Z}_2^{\times n}$ symmetries $U_1, \dots, U_n$. According to the result of~\cite{ShiozakiTopology2014}, each $\mathbb{Z}_2$ symmetry corresponds to four independent additional symmetry classes $t_i \in \{0,1,2,3\}$ for $i=1,\dots,n$. For each $t_i$, we use the binary representation $t_i = (t_{i1}t_{i0})_2$, i.e., 
\begin{align}
t_i = 2 t_{i1} + t_{i0}, \quad t_{i0}, t_{i1} \in \{0,1\}, \quad i=1,\dots,n.
\end{align}
Depending on the additional symmetry class $t_i$, the following relations hold~\cite{ShiozakiTopology2014}:
\begin{align}
\begin{cases}
U_i H(\bk,\br) U_i^{-1} = (-1)^{t_{i0}} H(P_i \bk, Q_i \br), \quad U_i^2 = 1, \\
A U_i = (-1)^{t_{i1} + (1-s_1)t_{i0}} U_i A, \\
\Gamma U_i = (-1)^{t_{i0}} U_i \Gamma \quad \text{(when $s_0=1$)}.
\end{cases}
\label{eq:add_sym_def}
\end{align}
Here $P_i$ and $Q_i$ represent diagonal $\mathbb{Z}_2$ actions on the variables $\bk$ and $\br$, respectively, defined by bit strings $p_i = (p_{i1},\dots,p_{id}) \in \{0,1\}^{\times d}$ and $q_i = (q_{i1},\dots,q_{iD}) \in \{0,1\}^{\times D}$:
\begin{align}
&P_i \bk = \left((-1)^{p_{i1}} k_1, \dots, (-1)^{p_{id}} k_d\right), \\
&Q_i \br = \left((-1)^{q_{i1}} r_1, \dots, (-1)^{q_{iD}} r_D\right).
\end{align}
See Table~\ref{Symmetry_type} for the correspondence with the notation used in~\cite{ShiozakiTopology2014}, which describes algebraic relations of additional symmetries.  
Within the sets of variables $\bk$ and $\br$, we introduce the numbers of variables simultaneously flipped by $U_i$:
\begin{align}
&d_{i_1 \cdots i_r} 
= \sum_{\mu=1}^d p_{i_1\mu} \cdots p_{i_r\mu}, \quad 1 \leq i_1 < \cdots < i_r \leq n, \\
&D_{i_1 \cdots i_r} 
= \sum_{\mu=1}^D q_{i_1\mu} \cdots q_{i_r\mu}, \quad 1 \leq i_1 < \cdots < i_r \leq n,
\end{align}
for $r = 1, \dots, n$.  
The algebraic relations among $U_i$ are labeled as
\begin{align}
U_i U_j = (-1)^{u_{ij}} U_j U_i, \quad u_{ij} \in \{0,1\}.
\end{align}

With these variables, the $K$-group can be expressed as
\begin{align}
K_{\mathbb{R}+nU}(s,\{t_i\}_i,\{u_{ij}\}_{i<j}; d,\{d_i\}_i,\{d_{ij}\}_{i<j},\{d_{ijk}\}_{i<j<k},\dots; D,\{D_i\}_i,\{D_{ij}\}_{i<j},\{D_{ijk}\}_{i<j<k},\dots),
\label{eq:Kgroup_def}
\end{align}
where $\mathbb{R}$ denotes the real AZ class and $nU$ indicates the presence of $n$ additional unitary $\mathbb{Z}_2$ symmetries. Following~\cite{TeoTopological2010}, we then construct the suspension isomorphism of $K$-theory, which raises the dimension of the sphere by one.

\begin{table}[!]
\begin{center}
\caption{
The relation between the real AZ classes, the additional symmetry classes, and the corresponding symmetry operators. The notation follows~\cite{ShiozakiTopology2014}. 
Here $U$ denotes a symmetry commuting with the Hamiltonian, while $\wb U$ denotes an antisymmetry anticommuting with the Hamiltonian. 
The superscript of $U$ indicates the sign of $U^2 = \pm 1$. For non-chiral classes ($s \equiv 0 \bmod 2$), the subscript indicates the commutation relation with TRS or PHS. For chiral classes ($s \equiv 1 \bmod 2$), the first and second subscripts indicate commutation/anticommutation with TRS and PHS, respectively.}
\label{Symmetry_type}
$$
\begin{array}{cc|cccc}
{\rm AZ\ class} & s \backslash t_i & 0 & 1 & 2 & 3 \\
\hline
{\rm AI}   & 0 &  U^+_+ & \wb U^+_- & U^+_- & \wb U^+_+ \\
{\rm BDI}  & 1 & U_{++}^+ & \wb U^+_{-+} & U^+_{--} & \wb U^+_{+-} \\
{\rm D}    & 2 & U^+_+ & \wb U^+_+ & U^+_- & \wb U^+_- \\
{\rm DIII} & 3 & U^+_{++} & \wb U^+_{-+} & U^+_{--} & \wb U^+_{+-} \\
{\rm AII}  & 4 & U^+_+ & \wb U^+_- & U^+_- & \wb U^+_+ \\
{\rm CII}  & 5 & U^+_{++} & \wb U^+_{-+} & U^+_{--} & \wb U^+_{+-} \\
{\rm C}    & 6 & U^+_+ & \wb U^+_+ & U^+_- & \wb U^+_- \\
{\rm CI}   & 7 & U^+_{++} & \wb U^+_{-+} & U^+_{--} & \wb U^+_{+-} \\
\end{array}
$$
\end{center}
\end{table}

\subsection{Non-chiral class $\rightarrow$ chiral class}

We construct the suspension isomorphism from a non-chiral class ($s_0=0$) to a chiral class ($s_0=1$). Let $\sigma_x,\sigma_y,\sigma_z$ be Pauli matrices, and define the Hamiltonian on the sphere $S^{d+D+1}$ as
\begin{align}
\tilde H(\bk,\br,\theta) = \cos \theta\, H(\bk,\br) \otimes \sigma_z + \sin \theta\, \sigma_x, \quad \theta \in [0,\pi].
\end{align}
At $\theta=0,\pi$, the dependence on $\bk,\br$ disappears, realizing the suspension $SS^{d+D} \cong S^{d+D+1}$. In what follows, the dependence on $\bk,\br$ is not needed, so we omit these variables. 

The Hamiltonian $\tilde H(\theta)$ has the chiral symmetry
\begin{align}
\sigma_y \tilde H(\theta) \sigma_y = - \tilde H(\theta).
\end{align}
The mappings of the TRS or PHS symmetry $A$ and the additional symmetries $U_i$ are not unique; they are specified by $\mathbb{Z}_2$ values $\delta s_1,\delta t_{10},\dots,\delta t_{n0} \in \{0,1\}$ within the freedom consistent with the definitions of symmetries in the chiral class~(\ref{eq:AZ_sym_chiral}), (\ref{eq:add_sym_def}). We define
\begin{align}
&\tilde A = A \otimes (\sigma_z)^{1-s_1}(\sigma_x)^{\delta s_1}, \quad \delta s_1 \in \{0,1\}, \\
&\tilde U_i = U_i \otimes (i)^{t_{i0}\delta t_{i0}} (\sigma_z)^{t_{i0}} (\sigma_x)^{\delta t_{i0}}, \quad \delta t_{i0} \in \{0,1\}, \quad i=1,\dots,n.
\end{align}
A straightforward computation yields the following symmetry relations and algebraic constraints:
\begin{align}
\begin{cases}
\tilde A \tilde H(\theta) \tilde A^{-1} = (-1)^{s_1+\delta s_1} \tilde H((-1)^{1-\delta s_1}\theta),\quad 
\tilde A^2 = (-1)^{s_2 + (1-s_1)\delta s_1}, \\
\tilde A \sigma_y = (-1)^{s_1+\delta s_1} \sigma_y \tilde A, \\
\tilde U_i \tilde H(\theta) \tilde U_i^{-1} = (-1)^{t_{i0}+\delta t_{i0}} \tilde H((-1)^{\delta t_{i0}}\theta),\quad \tilde U_i^2=1, \\
\tilde A \tilde U_i = (-1)^{t_{i1}+(t_{i0}+\delta s_1)\delta t_{i0} + (1-s_1+\delta s_1)(t_{i0}+\delta t_{i0})} \tilde U_i\tilde A, \\
\sigma_y \tilde U_i = (-1)^{t_{i0}+\delta t_{i0}} \tilde U_i \sigma_y, \\
\tilde U_i \tilde U_j = (-1)^{u_{ij} + t_{i0}\delta t_{j0}+ \delta t_{i0} t_{j0}} \tilde U_j \tilde U_i.
\end{cases}
\end{align}
From this, we obtain the mapping of the labels of symmetry classes:
\begin{align}
\begin{cases}
s_0=0 \mapsto s_0=1, \\
s_1 \mapsto s_1+\delta s_1, \\
s_2 \mapsto s_2+(1-s_1)\delta s_1, \\
t_{i0} \mapsto t_{i0}+\delta t_{i0}, \\
t_{i1} \mapsto t_{i1}+(t_{i0}+\delta s_1)\delta t_{i0}, \\
u_{ij} \mapsto u_{ij}+t_{i0}\delta t_{j0}+ \delta t_{i0} t_{j0}.
\end{cases}
\end{align}
Equivalently, when $s$ is even, this can be written as
\begin{align}
\begin{cases}
s \mapsto s + (-1)^{\delta s_1}, \\
t_i \mapsto t_i + \delta t_{i0} (-1)^{\delta s_1}, \\
u_{ij} \mapsto u_{ij}+t_{i0}\delta t_{j0}+ \delta t_{i0} t_{j0}.
\end{cases}
\label{eq:map_st}
\end{align}
The presence or absence of inversion of the variable $\theta$ is specified by $1-\delta s_1$ for $\tilde A$ and by $\delta t_{i0}$ for $\tilde U_i$. From the above, we obtain the following $K$-group isomorphism for even $s$:
\begin{align}
&K_{\mathbb{R}+nU}(s,\{t_i\}_i,\{u_{ij}\}_{i<j}; d,\{d_i\}_i,\{d_{ij}\}_{i<j},\{d_{ijk}\}_{i<j<k},\dots; D,\{D_i\}_i,\{D_{ij}\}_{i<j},\{D_{ijk}\}_{i<j<k}, \dots) \nonumber \\
&\xrightarrow{\cong}
K_{\mathbb{R}+nU}(s+1,\{t_i+\delta t_{i0}(-1)^{\delta s_1}\}_i,\{u_{ij}+t_{i0}\delta t_{j0}+ \delta t_{i0} t_{j0}\}_{i<j}; \nonumber \\
&\hspace{30pt} d+(1-\delta s_1),\{d_i + (1-\delta s_1)\delta t_{i0}\}_i,\{d_{ij} + (1-\delta s_1)\delta t_{i0}\delta t_{j0}\}_{i<j},\{d_{ijk}+(1-\delta s_1)\delta t_{i0}\delta t_{j0}\delta t_{k0}\}_{i<j<k},\dots;\nonumber\\
&\hspace{30pt} D+\delta s_1 ,\{D_i + \delta s_1\delta t_{i0}\}_i,\{D_{ij}+\delta s_1\delta t_{i0}\delta t_{j0}\}_{i<j},\{D_{ijk}+\delta s_1\delta t_{i0}\delta t_{j0}\delta t_{k0}\}_{i<j<k}, \dots).
\label{eq:iso}
\end{align}
As will be shown in the next subsection, this isomorphism also holds when $s$ is odd.

\subsection{Chiral class $\rightarrow$ non-chiral class}

In the same manner, we obtain a suspension isomorphism from the chiral class ($s_0=1$) to the non-chiral class ($s_0=0$). 
Define the Hamiltonian on $S^{d+D+1}$ by
\begin{align}
\tilde H(\theta) = \cos \theta\, H + \sin \theta\, \Gamma, \qquad \theta \in [0,\pi].
\end{align}
The mappings of the TRS or PHS operator $A$ and the additional symmetries $U_i$ are not unique; within the freedom compatible with the definitions of symmetries in the non-chiral class \eqref{eq:AZ_sym_nonchiral}, \eqref{eq:add_sym_def}, they are specified by $\mathbb{Z}_2$ values $\delta s_1,\delta t_{10},\dots,\delta t_{n0} \in \{0,1\}$. 
We set
\begin{align}
&\tilde A = \Gamma^{\,1-\delta s_1} A, \qquad \delta s_1 \in \{0,1\}, \\
&\tilde U_i = i^{\,t_{i0}\delta t_{i0}} \Gamma^{\,\delta t_{i0}} U_i, \qquad \delta t_{i0} \in \{0,1\}, \quad i=1,\dots,n.
\end{align}
A straightforward computation yields
\begin{align}
\begin{cases}
\tilde A \tilde H(\theta)\tilde A^{-1} = (-1)^{s_1+\delta s_1+1} \tilde H\!\left((-1)^{1-\delta s_1}\theta\right), \quad 
\tilde A^2 = (-1)^{\,s_2 + s_1(1-\delta s_1)}, \\
\tilde U_i \tilde H(\theta)\tilde U_i^{-1} = (-1)^{t_{i0}+\delta t_{i0}} \tilde H\!\left((-1)^{\delta t_{i0}}\theta\right), \quad \tilde U_i^2=1, \\
\tilde A \tilde U_i = (-1)^{t_{i1}+(t_{i0}+\delta s_1)\delta t_{i0} + (1-s_1+1-\delta s_1)(t_{i0}+\delta t_{i0})} \tilde U_i \tilde A, \\
\tilde U_i \tilde U_j = (-1)^{\,u_{ij}+ t_{i0}\delta t_{j0} + \delta t_{i0} t_{j0}} \tilde U_j \tilde U_i.
\end{cases}
\end{align}
Hence the labels map as
\begin{align}
\begin{cases}
s_0=1 \mapsto s_0=0, \\
s_1 \mapsto s_1 + (1-\delta s_1), \\
s_2 \mapsto s_2 + s_1(1-\delta s_1), \\
t_{i0} \mapsto t_{i0} + \delta t_{i0}, \\
t_{i1} \mapsto t_{i1} + (t_{i0}+\delta s_1)\delta t_{i0}, \\
u_{ij} \mapsto u_{ij} + t_{i0}\delta t_{j0} + \delta t_{i0} t_{j0}.
\end{cases}
\end{align}
This is equivalent to \eqref{eq:map_st} for even $s$. Therefore, we obtain the $K$-group isomorphism \eqref{eq:iso} for odd $s$ as well.

\subsection{Reduction to zero dimension}

From the isomorphism \eqref{eq:iso}, the $K$-group reduces to that in zero dimension, i.e.\ $d=D=0$\footnote{%
To derive \eqref{eq:iso_to_zero}, it is convenient to use the transformation rule for the suspended variables $\tilde u_{ij}:=u_{ij}+t_i t_j$, namely $\tilde u_{ij}\mapsto \tilde u_{ij} + \delta t_{i0}\delta t_{j0}$ following \eqref{eq:map_st}. 
Applying \eqref{eq:iso} a total of $d+D$ times, the symmetry class in $0$D changes as $(\{t_i\}_i,\{\tilde u_{ij}\}_{i<j}) \mapsto (\{t_i-\delta_i\}_i,\{\tilde u_{ij}-\delta_{ij}\}_{i<j})$. 
Since $u_{ij}=\tilde u_{ij}-t_i t_j$, we get $u_{ij}\mapsto \tilde u_{ij}-\delta_{ij}-(t_i-\delta_i)(t_j-\delta_j)$, which yields \eqref{eq:iso_to_zero}.}:
\begin{align}
&K_{\mathbb{R}+nU}(s,\{t_i\}_i,\{u_{ij}\}_{i<j};\, d,\{d_i\}_i,\{d_{ij}\}_{i<j},\{d_{ijk}\}_{i<j<k},\dots;\, D,\{D_i\}_i,\{D_{ij}\}_{i<j},\{D_{ijk}\}_{i<j<k},\dots) \nonumber\\
&\qquad \cong\;
K_{\mathbb{R}+nU}\bigl(s-\delta,\{t_i-\delta_i\}_i,\{u_{ij}+\delta_{ij}+t_i\delta_j+t_j\delta_i+\delta_i \delta_j\}_{i<j}\bigr).
\label{eq:iso_to_zero}
\end{align}
Here we introduced
\begin{align}
\delta = d - D, \qquad 
\delta_i = d_i - D_i, \qquad 
\delta_{ij} = d_{ij} - D_{ij}.
\end{align}
In the zero-dimensional case as on the right-hand side of \eqref{eq:iso_to_zero}, we will omit $d,D$ from the notation. 
As in the cases with only internal symmetries~\cite{TeoTopological2010} and with a single $\mathbb{Z}_2$ point-group symmetry~\cite{ShiozakiTopology2014}, the $K$-group depends only on the defect dimension $(\delta-1)$ and on the labels $\delta_i,\delta_{ij}$ that characterize how the $\mathbb{Z}_2^{\times n}$ point-group symmetry acts on the $\delta$-dimensional real space whose boundary is the defect.

Consequently, the $K$-group does not depend on the numbers of variables flipped in common by three or more generators, i.e.\ on $d_{ijk}, d_{ijkl},\dots$ and $D_{ijk}, D_{ijkl},\dots$. 
This follows from the fact that, in the isomorphism \eqref{eq:iso}, the labels specifying the symmetry class $s,t_i,u_{ij}$ are insensitive to the product $\delta t_{i_10}\cdots \delta t_{i_r0}$ with $r\ge 3$ indicating triple or higher overlaps. 
A minimal example exhibiting independence of $d_{ijk}$ is $d=4$, $d_1=d_2=d_3=2$, $D=0$, for which there are two possibilities $d_{123}=0,1$ but they give the same $K$-group.

In this paper, we do not explicitly compute the $K$-groups and periodic tables for three or more additional $\mathbb{Z}_2$ point-group symmetries. 
The computation is straightforward and proceeds in the same manner as in Sec.~\ref{sec:Kgroup_compute}.

\subsection{Complex AZ classes with unitary additional symmetries}

For complex AZ classes (A, AIII) in the presence of additional $\mathbb{Z}_2^{\times n}$ point-group symmetries, the $K$-group isomorphism \eqref{eq:iso_to_zero} also holds. 
In this setting, both the label $s$ specifying the complex AZ class and the labels $t_i$ specifying the additional symmetry types have period~2, i.e., $s,t_i \in \{0,1\}$. 
Table~\ref{Symmetry_type_complex} summarizes the relation between $(s,t_i)$ and the symmetry operators.

\begin{table}[!]
\begin{center}
\caption{
Complex AZ classes, additional symmetry classes, and the corresponding symmetry operators. 
The notation follows~\cite{ShiozakiTopology2014}. 
Here $U$ denotes a symmetry commuting with the Hamiltonian, while $\wb U$ denotes an antisymmetry anticommuting with the Hamiltonian. 
For the chiral class ($s=1$), the subscript of $U$ indicates commutation/anticommutation with the chiral operator. 
Since there is no antiunitary symmetry, one can always redefine $U$ so that $U^2=1$, hence superscripts are omitted.
}
\label{Symmetry_type_complex}
$$
\begin{array}{cc|cc}
{\rm AZ\ class} & s \backslash t_i & 0 & 1\\
\hline
{\rm A}    & 0 & U   & \wb U \\
{\rm AIII} & 1 & U_+ & \wb U_- \\
\end{array}
$$
\end{center}
\end{table}

\subsection{Complex AZ classes with antiunitary additional symmetries}

For complex AZ classes (A, AIII) with additional antiunitary $\mathbb{Z}_2^{\times n}$ point-group symmetries, the problem reduces to the case of real AZ classes with $\mathbb{Z}_2^{\times (n-1)}$ additional symmetries, in the same manner as in~\cite{ShiozakiTopology2014}.

To avoid cumbersome notation, we present the case $n=2$ for clarity. 
Since the product of two antiunitary symmetries is unitary, without loss of generality we take the first generator of $\mathbb{Z}_2^{\times n}$ to be antiunitary and the second to be unitary. 
We denote the antiunitary symmetry by $A$ and the unitary one by $U_1$.
Combining the complex AZ class with the antiunitary symmetry $A$ yields eight symmetry classes, which we label by $s \in \{0,\dots,7\}$ with the binary representation $s=(s_2 s_1 s_0)_2$, in analogy with \eqref{eq:s_2adic}, \eqref{eq:AZ_sym_nonchiral}, and \eqref{eq:AZ_sym_chiral}.
Let $P_A=\mathrm{diag}((-1)^{p_{A,1}},\dots,(-1)^{p_{A,d}})$ be the $\mathbb{Z}_2$ action of $A$ on momentum coordinates and $Q_A=\mathrm{diag}((-1)^{q_{A,1}},\dots,(-1)^{q_{A,D}})$ that on the defect (real-space–like) parameters. 
Taking into account the momentum inversion due to the antiunitarity of $A$, the symmetries of $H(\bk,\br)$ read
\begin{align}
\text{$s_0=0$ (non-chiral class)}:\quad
A H(\bk,\br) A^{-1} = (-1)^{s_1} H(-P_A \bk, Q_A \br),\qquad A^2 = (-1)^{s_2},
\label{eq:sym_additional_A_nc}
\end{align}
\begin{align}
\text{$s_0=1$ (chiral class)}:\quad
\begin{cases}
A H(\bk,\br) A^{-1} = (-1)^{s_1} H(-P_A \bk, Q_A \br),\qquad A^2 = (-1)^{s_2},\\
\Gamma H(\bk,\br) \Gamma^{-1} = - H(\bk,\br),\qquad \Gamma^2=1,\\
A \Gamma = (-1)^{s_1} \Gamma A.
\end{cases}
\label{eq:sym_additional_A_c}
\end{align}
For the additional unitary symmetry $U_1$ we introduce an additional class $t_1\in\{0,1,2,3\}$ and use the binary representation $t_1=(t_{11} t_{10})_2$ to specify the algebraic relations, in analogy with \eqref{eq:add_sym_def}.

Define the numbers of momenta flipped by $A$ and $U_1$ as
\begin{align}
d_A = \sum_{\mu=1}^d p_{A,\mu},\qquad 
d_1 = \sum_{\mu=1}^d p_{1,\mu},
\end{align}
and similarly the numbers of defect parameters flipped by $A$ and $U_1$ as
\begin{align}
D_A = \sum_{\mu=1}^D q_{A,\mu},\qquad 
D_1 = \sum_{\mu=1}^D q_{1,\mu}.
\end{align}
The numbers of variables flipped simultaneously by $A$ and $U_1$ are
\begin{align}
d_{A1} = \sum_{\mu=1}^d p_{A,\mu} p_{1,\mu},\qquad 
D_{A1} = \sum_{\mu=1}^D q_{A,\mu} q_{1,\mu}.
\end{align}
We denote the $K$-group by
\begin{align}
K_{\mathbb{C}+A+U}(s,t;\, d,d_A,d_1,d_{A1};\, D,D_A,D_1,D_{A1}).
\end{align}

Now, the symmetry of $A$ in \eqref{eq:sym_additional_A_nc}, \eqref{eq:sym_additional_A_c} can be viewed as TRS or PHS of a real AZ class if we define the effective numbers of momentum-like and real-space–like variables flipped by $A$ as
\begin{align}
&\tilde d = \sum_{\mu=1}^d (1-p_{A,\mu}) + \sum_{\mu=1}^D q_{A,\mu} = d - d_A + D_A,\\
&\tilde D = \sum_{\mu=1}^D (1-q_{A,\mu}) + \sum_{\mu=1}^d p_{A,\mu} = D - D_A + d_A,
\end{align}
as explained in~\cite{ShiozakiTopology2014}. 
With this effective partition $(\tilde d,\tilde D)$, the numbers of variables flipped by $U_1$ become
\begin{align}
&\tilde d_1 = \sum_{\mu=1}^d (1-p_{A,\mu}) p_{1,\mu} + \sum_{\mu=1}^D q_{A,\mu} q_{1,\mu} = d_1 - d_{A1} + D_{A1},\\
&\tilde D_1 = \sum_{\mu=1}^D (1-q_{A,\mu}) q_{1,\mu} + \sum_{\mu=1}^d p_{A,\mu} p_{1,\mu} = D_1 - D_{A1} + d_{A1}.
\end{align}
Therefore we obtain the isomorphisms
\begin{align}
K_{\mathbb{C}+A+U}(s,t;\, d,d_A,d_1,d_{A1};\, D,D_A,D_1,D_{A1})
&\cong K_{\mathbb{R}+U}(s,t;\, \tilde d,\tilde d_1;\, \tilde D,\tilde D_1)\nonumber \\
&\cong K_{\mathbb{R}+U}(s-\tilde d + \tilde D,\; t-\tilde d_1+\tilde D_1)\nonumber \\
&= K_{\mathbb{R}+U}(s-\delta + 2\delta_A,\; t-\delta_1 + 2\delta_{A1}),
\end{align}
where we set $\delta = d-D$, $\delta_A = d_A - D_A$, $\delta_1 = d_1 - D_1$, and $\delta_{A1} = d_{A1} - D_{A1}$.
The groups $K_{\mathbb{R}+U}(s,t)$ are computed in~\cite{ShiozakiTopology2014} (see Table~\ref{tab:AZ+U} below).

\subsection{$K$-groups on tori}

The isomorphism \eqref{eq:iso_to_zero} holds when the $(d+D)$-dimensional parameter space is a sphere. 
For a torus, contributions from all sub-tori appear as direct summands.

For example, when the momentum space is a two-dimensional torus $T^2$ and the defect parameter space is a one-dimensional circle $S^1$, the total space is $T^2 \times S^1$. 
Denoting the coordinates by $k_1,k_2,r_1$, the following eight subspaces contribute as direct-sum components:
\begin{align}
\begin{cases}
S^3: (k_1,k_2,r_1),\\
S^2: (k_1,k_2),\\
S^2: (k_1,r_1),\\
S^2: (k_2,r_1),\\
S^1: (k_1),\\
S^1: (k_2),\\
S^1: (r_1),\\
S^0.
\end{cases}
\end{align}
On each subspace, the $K$-group \eqref{eq:Kgroup_def} is defined according to the numbers of variables flipped by the additional symmetries $U_i$, and the explicit classification groups are determined via the isomorphism \eqref{eq:iso_to_zero}.

\section{Classification tables with additional $\mathbb{Z}_2^{\times 2}$ point group symmetry}
In this section, the $0$-dimensional $K$-groups are explicitly computed for the case with additional $\mathbb{Z}_2^{\times 2}$ point group symmetry, and the results are summarized in classification tables. 
For comparison, the case with a single additional $\mathbb{Z}_2$ symmetry was calculated in~\cite{ShiozakiTopology2014}, and the corresponding results are collected in Table~\ref{tab:AZ+U}.

\begin{table}[!]
\begin{center}
\caption{Classifying spaces in the presence of a single additional $\mathbb{Z}_2$ point-group symmetry of class $t$, in addition to a complex or real AZ class labeled by $s$~\cite{ShiozakiTopology2014}.}
$$
\begin{array}{c|c|c}
\text{Complex/Real} & \text{Additional symmetry class } t & \text{Classifying space} \\
\hline
\text{Complex} & 0 & (C_s)^{\times 2} \\
               & 1 & C_{s+1} \\
\hline
\multirow{4}{*}{\text{Real}} & 0 & (R_s)^{\times 2} \\
                             & 1 & R_{s-1} \\
                             & 2 & C_s \\
                             & 3 & R_{s+1} \\
\end{array}
$$
\label{tab:AZ+U}
\end{center}
\end{table}

\subsection{Computation of effective AZ classes in zero dimension}
\label{sec:Kgroup_compute}

Due to the hierarchical structure, it is sufficient to determine the AZ classes for $s=0$; the classifying spaces for arbitrary $s$ are then obtained by shifting the index accordingly. 
The procedure for computing AZ classes has already been established in~\cite{ShiozakiAtiyahHirzebruch2022}. 
In that approach, one first decomposes the unitary symmetry group into its irreducible representations and then applies the Wigner criteria, which provides a systematic calculation. 
Here, we instead perform the computation explicitly by a more elementary hand calculation.

For each set of additional symmetry classes $(t_1,t_2,u_{12})$, we introduce the following convenient notation. 
We denote the two unitary $\mathbb{Z}_2$ symmetries by $U_1$ and $U_2$. 
As in Table~\ref{Symmetry_type}, we specify the signs of $(U_1)^2$ and $(U_2)^2$, their commutation relations with TRS and/or PHS, and whether they act as symmetries $U$ (commuting with the Hamiltonian) or antisymmetries $\wb U$ (anticommuting with the Hamiltonian). 
The mutual relation is encoded in the sign of $U_1 U_2 = (-1)^{u_{12}} U_2 U_1$ so that the algebraic data of the additional symmetries can be summarized by the triple
\begin{align}
(U_1,\,U_2,\,(-1)^{u_{12}}).
\end{align}
For example, in the case $(t_1,t_2,u_{12})=(1,2,0)$, the notation reads $(\wb U_1{}^+_-,\, U_2{}^+_-,\,+)$.

We now compute the AZ classes for $s=0$. 
Note that the AZ class is invariant under the redefinitions
\begin{align}
(U_1,U_2)\;\mapsto\; (U_2,U_1),\qquad (U_1,\, (i)^{u_{12}} U_1U_2),\qquad ((i)^{u_{12}} U_1U_2,\,U_2).
\end{align}
These redefinitions generate equivalences among the symmetry labels,
\begin{align}
(t_1,t_2,u_{12}) \;\sim\; (t_2,t_1,u_{12}) \;\sim\; (t_1,\,t_2+t_1+2t_1t_2+2u_{12},\,u_{12}) \;\sim\; (t_1+t_2+2t_1t_2+2u_{12},\,t_2,\,u_{12}).
\end{align}
Hence, in the following, it is sufficient to restrict to the case $t_1 \leq t_2$.

\medskip

\noindent
Complex AZ classes with additional unitary $\mathbb{Z}_2^{\times 2}$ symmetries---
\begin{itemize}
\item
$(t_1,t_2,u_{12})=(0,0,0): (U_1,U_2,+)$.

The system splits into four one-dimensional sectors labeled by $U_1=\pm1$, $U_2=\pm1$. 
Thus, $({\rm A})^{\times 4}$.

\item
$(t_1,t_2,u_{12})=(0,1,0): (U_1,\wb{U}_2,+)$, with $(0,1,0)\sim(1,1,0)$.

The system decomposes into blocks with $U_1=\pm1$. 
Since $U_1U_2=U_2U_1$, the operator $U_2$ acts as a chiral symmetry in each sector. 
Thus, $({\rm AIII})^{\times 2}$.

\item
$(t_1,t_2,u_{12})=(0,0,1): (U_1,U_2,-)$.

Since $U_1U_2=-U_2U_1$, the pair forms a nontrivial two-dimensional projective representation of $\mathbb{Z}_2\times\mathbb{Z}_2$. 
Thus, ${\rm A}$.

\item
$(t_1,t_2,u_{12})=(0,1,1): (U_1,\wb{U}_2,-)$, with $(0,1,1)\sim(1,1,1)$.

Because $U_1U_2=-U_2U_1$, the operator $U_2$ exchanges the sectors $U_1=\pm1$ and acts as an antisymmetry. 
Thus, ${\rm A}$.
\end{itemize}

\medskip

\noindent
Real AZ classes with additional unitary $\mathbb{Z}_2^{\times 2}$ symmetries---
\begin{itemize}
\item
$(t_1,t_2,u_{12})=(0,0,0): (U_1{}^+_+,U_2{}^+_+,+)$.

In the four one-dimensional sectors $U_1=\pm1$, $U_2=\pm1$, TRS $T$ closes. 
Thus, $({\rm AI})^{\times 4}$.

\item
$(t_1,t_2,u_{12})=(0,1,0): (U_1{}^+_+,\wb{U}_2{}^+_-,+)$, with $(0,1,0)\sim(1,1,0)$.

Since $U_1$ commutes with both $U_2$ and $T$, the system decomposes into blocks with $U_1=\pm1$. 
In each sector, TRS with $T^2=1$ exists, together with a PHS $(U_2T)$ satisfying $(U_2T)^2=-1$. 
Thus, $({\rm CI})^{\times 2}$.

\item
$(t_1,t_2,u_{12})=(0,2,0): (U_1{}^+_+,U_2{}^+_-,+)$, with $(0,2,0)\sim(2,2,0)$.

Since $U_1$ commutes with $U_2$ and $T$, the system splits into blocks with $U_1=\pm1$. 
Because $TU_2=-U_2T$, TRS $T$ exchanges the sectors $U_2=\pm1$. 
Thus, $({\rm A})^{\times 2}$.

\item
$(t_1,t_2,u_{12})=(0,3,0): (U_1{}^+_+,\wb{U}_2{}^+_+,+)$, with $(0,3,0)\sim(3,3,0)$.

Since $U_1$ commutes with $U_2$ and $T$, the system splits into blocks with $U_1=\pm1$. 
In each sector, TRS with $T^2=1$ and PHS with $(U_2T)^2=1$ are present. 
Thus, $({\rm BDI})^{\times 2}$.

\item
$(t_1,t_2,u_{12})=(1,2,0): (\wb{U}_1{}^+_-,U_2{}^+_-,+)$, with $(1,2,0)\sim(1,3,0)\sim(2,3,0)$.

Since $TU_2=-U_2T$, TRS $T$ exchanges the sectors $U_2=\pm1$. 
Because $U_1U_2=U_2U_1$, the operator $U_1$ acts as a chiral symmetry within each $U_2$ sector. 
Thus, ${\rm AIII}$.

\item
$(t_1,t_2,u_{12})=(0,0,1): (U_1{}^+_+,U_2{}^+_+,-)$, with $(0,0,1)\sim(0,2,1)$.

Since $TU_1=U_1T$, TRS $T$ closes in the sectors $U_1=\pm1$. 
Because $U_1U_2=-U_2U_1$, the operator $U_2$ exchanges the sectors $U_1=\pm1$. 
Thus, ${\rm AI}$.

\item
$(t_1,t_2,u_{12})=(0,1,1): (U_1{}^+_+,\wb{U}_2{}^+_-,-)$, with $(0,1,1)\sim(0,3,1)\sim(1,3,1)$.

Since $TU_1=U_1T$, TRS $T$ closes in the sectors $U_1=\pm1$. 
Because $U_1U_2=-U_2U_1$, the operator $U_2$ exchanges the sectors $U_1=\pm1$ as an antisymmetry. 
Thus, ${\rm AI}$.

\item
$(t_1,t_2,u_{12})=(1,2,1): (\wb{U}_1{}^+_-,U_2{}^+_-,-)$, with $(1,2,1)\sim(1,1,1)$.

Since $TU_2=-U_2T$, TRS $T$ exchanges the sectors $U_2=\pm1$. 
Because $U_1TU_2=U_2U_1T$, the operator $U_1T$ closes within each $U_2$ sector and acts as a PHS with $(U_1T)^2=-1$. 
Thus, ${\rm C}$.

\item
$(t_1,t_2,u_{12})=(2,2,1): (U_1{}^+_-,U_2{}^+_-,-)$.

Since $TU_1=-U_1T$, TRS $T$ exchanges the sectors $U_1=\pm1$. 
Because $U_2TU_1=U_1U_2T$, the operator $U_2T$ acts as a TRS with $(U_2T)^2=-1$. 
Thus, ${\rm AII}$.

\item
$(t_1,t_2,u_{12})=(2,3,1): (U_1{}^+_-,\wb{U}_2{}^+_+,-)$, with $(2,3,1)\sim(3,3,1)$.

Since $TU_1=-U_1T$, TRS $T$ exchanges the sectors $U_1=\pm1$. 
Because $U_2TU_1=U_1U_2T$, the operator $U_2T$ closes within each $U_1$ sector and acts as a PHS with $(U_2T)^2=1$. 
Thus, ${\rm D}$.
\end{itemize}

From the above analysis, we have obtained the effective AZ classes realized for each additional symmetry class at $s=0$. 
For a general $s$, the classifying space is obtained by shifting the AZ index by $s$, that is, by taking $s+s_0$ where $s_0$ corresponds to the effective AZ class. 
The results are summarized in Table~\ref{tab:realAZ+U1U2_classifying_space}.

\begin{table}[!]
\begin{center}
\caption{Classifying spaces in the presence of additional $\mathbb{Z}_2^{\times 2}$ symmetries specified by $(t_1,t_2,u_{12})$, on top of an AZ class labeled by $s$. 
Since the results are invariant under exchanging $t_1 \leftrightarrow t_2$, only the cases with $t_1 \leq t_2$ are shown. }
$$
\begin{array}{ccccccccccccc}
{\rm Complex/Real}&{\rm Additional\ symmetry\ class\ }(t_1,t_2,u_{12})&{\rm AZ\ class\ for\ }s=0&{\rm Classifying\ space}\\
\hline
&(0,0,0)&({\rm A})^{\times 4}&(C_s)^{\times 4}\\
{\rm Complex}&(0,1,0) \sim (1,1,0)&({\rm AIII})^{\times 2}&(C_{s+1})^{\times 2}\\
&(0,0,1), (0,1,1) \sim (1,1,1)&{\rm A}&C_s\\
\hline
&(0,0,0)&({\rm AI})^{\times 4}&(R_s)^{\times 4}\\
&(0,1,0) \sim (1,1,0)&({\rm CI})^{\times 2}&(R_{s-1})^{\times 2}\\
&(0,2,0) \sim (2,2,0)&({\rm A})^{\times 2}&(C_s)^{\times 2}\\
&(0,3,0) \sim (3,3,0)&({\rm BDI})^{\times 2}&(R_{s+1})^{\times 2}\\
{\rm Real}&(1,2,0) \sim (1,3,0) \sim (2,3,0)&{\rm AIII}&C_{s+1}\\
&(0,0,1) \sim (0,2,1), (0,1,1) \sim (0,3,1) \sim (1,3,1)&{\rm AI}&R_{s}\\
&(1,1,1) \sim (1,2,1)&{\rm C}&R_{s-2}\\
&(2,2,1)&{\rm AII}&R_{s+4}\\
&(2,3,1) \sim (3,3,1)&{\rm D}&R_{s+2}\\
\end{array}
$$
\label{tab:realAZ+U1U2_classifying_space}
\end{center}
\end{table}

The Abelian groups indicating the number of connected components of each classifying space (denoted by $\pi_0$) are listed in Table~\ref{tab:classifying_space}. 
From these results, the complete classification table is obtained for the case of real AZ classes with two additional unitary $\mathbb{Z}_2$ symmetries. 
In the following sections, we classify the cases according to the number of variables flipped by $U_1$ and $U_2$, and present the explicit classification tables.

\begin{table}[!]
\begin{center}
\caption{AZ classes and classifying spaces. 
The periodicity of $s$ is $2$ for complex AZ classes and $8$ for real AZ classes. 
$\pi_0$ denotes the number of connected components.}
$$
\begin{array}{cccccccc}
{\rm AZ\ class}&s&{\rm Classifying\ space}&\pi_0\\
\hline \
{\rm A}&0&C_0&\Z\\
{\rm AIII}&1&C_1&0\\
{\rm AI}&0&R_0&\Z\\
{\rm BDI}&1&R_1&\Z_2\\
{\rm D}&2&R_2&\Z_2\\
{\rm DIII}&3&R_3&0\\
{\rm AII}&4&R_4&\Z\\
{\rm CII}&5&R_5&0\\
{\rm C}&6&R_6&0\\
{\rm CI}&7&R_7&0\\
\end{array}
$$
\label{tab:classifying_space}
\end{center}
\end{table}

\subsection{Periodic tables for additional unitary $\mathbb{Z}_2^{\times 2}$ point-group symmetry}

In this subsection, we present classification tables for the case where, in addition to complex or real AZ symmetries, an additional unitary $\mathbb{Z}_2^{\times 2}$ point-group symmetry is present. 
The $K$-group isomorphism is given by
\begin{align}
&K_{\mathbb{C}/\mathbb{R}+2U}(s,t_1,t_2,u_{12};d,d_1,d_2,d_{12};D,D_1,D_2,D_{12}) \nonumber \\ 
&\cong
K_{\mathbb{C}/\mathbb{R}+2U}(s-\delta,\,t_1-\delta_1,\,t_2-\delta_2,\,u_{12}+\delta_{12}+t_1\delta_2+t_2\delta_1+\delta_1\delta_2).
\label{eq:iso_zero_2U}
\end{align}
Although in principle all possibilities of $(\delta_1,\delta_2,\delta_{12})$ may be considered, only the seven cases listed below are physically realized in spatial dimensions up to three.
\begin{itemize}
\item 
$\Z_2$ onsite + $\Z_2$ onsite.

This is the case where $\delta_1=\delta_2=0$. 
\item 
$\Z_2$ onsite + reflection. 

This is the case where $\delta_1=0, \delta_2=1$. 
\item 
$\Z_2$ onsite + $C_2$ rotation.

This is the case where $\delta_1=0, \delta_2=2$. 
\item 
$\Z_2$ onsite + inversion. 

This is the case where $\delta_1=0, \delta_2=3$. 
\item 
Reflection + reflection (with a different reflection plane). 

This is the case where $\delta_1=1, \delta_2=1$, and $\delta_{12}=0$. 
Note that the two reflections have different reflection planes; otherwise, the redefinition of $U_1$ as $U_1U_2$ is recast as a $\Z_2$ on-site and reflection.
This case is possible when the defect dimensions are larger than two.
\item 
Reflection + $C_2$ rotation where the reflection plane is perpendicular to the rotation axis. 

This is the case where $\delta_1=1, \delta_2=2$, and $\delta_{12}=0$. 
Note that when the rotation axis is parallel to the reflection plane ($\delta_{12}=1$), then the redefinition of $U_2$ as $U_1U_2$ is recast as two reflections with $\delta_{12}=0$. 
This is only possible for bulk classification in three dimensions. 
\item 
$C_2$ rotation + $C_2$ rotation (with the rotation axes perpendicular to each other). 

This is the case where $\delta_1=2, \delta_2=2$, and $\delta_{12}=1$. 
This is only possible for bulk classification in three dimensions. 
\end{itemize}

The classification tables depend only on the four parameters obtained from the isomorphism~\eqref{eq:iso_zero_2U},
\begin{align}
    \tilde s=s-\delta,\quad
    \tilde t_1=t_1-\delta_1,\quad
    \tilde t_2=t_2-\delta_2,\quad
    \tilde u_{12}=u_{12}+\delta_{12}+t_1\delta_2+t_2\delta_1+\delta_1\delta_2.
\end{align}
Therefore, it is sufficient to provide classification tables for the twelve cases of $(\tilde t_1,\tilde t_2,\tilde u_{12})$ listed in Table~\ref{tab:realAZ+U1U2_classifying_space}. 
The results are summarized in Table~\ref{tab:table_2U}.

\newgeometry{left=2cm,right=2cm,top=2cm,bottom=2cm}
\begin{table}[!]
\begin{center}
\caption{
Periodic table for unitary $\mathbb{Z}_2^{\times 2}$ point-group symmetry. 
Here, $\delta_1$ and $\delta_2$ denote the numbers of variables flipped by $U_1$ and $U_2$, respectively, while $\delta_{12}$ denotes the number of variables flipped simultaneously by both $U_1$ and $U_2$. 
For each AZ class, the additional symmetry classes $t_1,t_2$ associated with $U_1,U_2$ are determined by Tables~\ref{Symmetry_type} and~\ref{Symmetry_type_complex}. 
The parameter $u_{12}\in\{0,1\}$ specifies whether $U_1$ and $U_2$ commute or anticommute. 
Given the data $\delta_1,\delta_2,\delta_{12},t_1,t_2,u_{12}$ of the unitary $\mathbb{Z}_2^{\times 2}$ point-group symmetry, one computes the triple $(t_1-\delta_1,\,t_2-\delta_2,\,u_{12}+\delta_{12}+t_1\delta_2+t_2\delta_1+\delta_1\delta_2)$, 
and the corresponding classification result is obtained from the relevant row in the first column.
}
\resizebox{\textwidth}{!}{
\begin{tabular}[t]{ccccccccccc}
$(t_1-\delta_1,t_2-\delta_2,u_{12}+\delta_{12}+t_1\delta_2+t_2\delta_1+\delta_1\delta_2)$ & Classifying space & AZ class 
& $\delta=0$ & $\delta=1$ & $\delta=2$ & $\delta=3$ 
& $\delta=4$ & $\delta=5$ & $\delta=6$ & $\delta=7$ \\
\hline
\multirow{2}{*}{$(0,0,0)$}
&\multirow{2}{*}{$(C_{s-\delta})^4$}
&A& $\mathbb{Z}^4$ & $0$ & $\mathbb{Z}^4$ & $0$ & $\mathbb{Z}^4$ & $0$ & $\mathbb{Z}^4$ & $0$ \\ 
&&AIII&$0$ & $\mathbb{Z}^4$ & $0$ & $\mathbb{Z}^4$ & $0$ & $\mathbb{Z}^4$ & $0$ & $\mathbb{Z}^4$ \\ 
\hline
\multirow{2}{*}{$(1,0,0)\sim (0,1,0)\sim (1,1,0)$}
&\multirow{2}{*}{$(C_{s+1-\delta})^2$}
& A & $0$ & $\mathbb{Z}^2$ & $0$ & $\mathbb{Z}^2$ & $0$ & $\mathbb{Z}^2$ & $0$ & $\mathbb{Z}^2$ \\ 
&& AIII & $\mathbb{Z}^2$ & $0$ & $\mathbb{Z}^2$ & $0$ & $\mathbb{Z}^2$ & $0$ & $\mathbb{Z}^2$ & $0$ \\ 
\hline
\multirow{2}{*}{
$\begin{array}{cc}
(0,0,1),\\
(1,0,1)\sim (0,1,1)\sim (1,1,1)\\
\end{array}$}
&\multirow{2}{*}{$C_{s-\delta}$}
& A & $\mathbb{Z}$ & $0$ & $\mathbb{Z}$ & $0$ & $\mathbb{Z}$ & $0$ & $\mathbb{Z}$ & $0$ \\ 
&& AIII & $0$ & $\mathbb{Z}$ & $0$ & $\mathbb{Z}$ & $0$ & $\mathbb{Z}$ & $0$ & $\mathbb{Z}$ \\ 
\hline
\multirow{8}{*}{$(0,0,0)$}
&\multirow{8}{*}{$(R_{s-\delta})^4$}
& AI   &$\mathbb{Z}^4$ & $0$ & $0$ & $0$ & $(2\mathbb{Z})^4$ & $0$ & $\mathbb{Z}_2^4$ & $\mathbb{Z}_2^4$ \\
&& BDI &$\mathbb{Z}_2^4$ & $\mathbb{Z}^4$ & $0$ & $0$ & $0$ & $(2\mathbb{Z})^4$ & $0$ & $\mathbb{Z}_2^4$ \\
&& D    &$\mathbb{Z}_2^4$ & $\mathbb{Z}_2^4$ & $\mathbb{Z}^4$ & $0$ & $0$ & $0$ & $(2\mathbb{Z})^4$ & $0$ \\
&& DIII &$0$ & $\mathbb{Z}_2^4$ & $\mathbb{Z}_2^4$ & $\mathbb{Z}^4$ & $0$ & $0$ & $0$ & $(2\mathbb{Z})^4$ \\
&& AII  &$(2\mathbb{Z})^4$ & $0$ & $\mathbb{Z}_2^4$ & $\mathbb{Z}_2^4$ & $\mathbb{Z}^4$ & $0$ & $0$ & $0$ \\
&& CII  &$0$ & $(2\mathbb{Z})^4$ & $0$ & $\mathbb{Z}_2^4$ & $\mathbb{Z}_2^4$ & $\mathbb{Z}^4$ & $0$ & $0$ \\
&& C    &$0$ & $0$ & $(2\mathbb{Z})^4$ & $0$ & $\mathbb{Z}_2^4$ & $\mathbb{Z}_2^4$ & $\mathbb{Z}^4$ & $0$ \\
&& CI   &$0$ & $0$ & $0$ & $(2\mathbb{Z})^4$ & $0$ & $\mathbb{Z}_2^4$ & $\mathbb{Z}_2^4$ & $\mathbb{Z}^4$ \\
\hline
\multirow{8}{*}{$(1,0,0)\sim(0,1,0)\sim(1,1,0)$}
&\multirow{8}{*}{$(R_{s-1-\delta})^2$}
& AI  & $0$ & $0$ & $0$ & $(2\mathbb{Z})^2$ & $0$ & $\mathbb{Z}_2^2$ & $\mathbb{Z}_2^2$ & $\mathbb{Z}^2$ \\
&& BDI & $\mathbb{Z}^2$ & $0$ & $0$ & $0$ & $(2\mathbb{Z})^2$ & $0$ & $\mathbb{Z}_2^2$ & $\mathbb{Z}_2^2$ \\
&& D   & $\mathbb{Z}_2^2$ & $\mathbb{Z}^2$ & $0$ & $0$ & $0$ & $(2\mathbb{Z})^2$ & $0$ & $\mathbb{Z}_2^2$ \\
&& DIII& $\mathbb{Z}_2^2$ & $\mathbb{Z}_2^2$ & $\mathbb{Z}^2$ & $0$ & $0$ & $0$ & $(2\mathbb{Z})^2$ & $0$ \\
&& AII & $0$ & $\mathbb{Z}_2^2$ & $\mathbb{Z}_2^2$ & $\mathbb{Z}^2$ & $0$ & $0$ & $0$ & $(2\mathbb{Z})^2$ \\
&& CII & $(2\mathbb{Z})^2$ & $0$ & $\mathbb{Z}_2^2$ & $\mathbb{Z}_2^2$ & $\mathbb{Z}^2$ & $0$ & $0$ & $0$ \\
&& C   & $0$ & $(2\mathbb{Z})^2$ & $0$ & $\mathbb{Z}_2^2$ & $\mathbb{Z}_2^2$ & $\mathbb{Z}^2$ & $0$ & $0$ \\
&& CI  & $0$ & $0$ & $(2\mathbb{Z})^2$ & $0$ & $\mathbb{Z}_2^2$ & $\mathbb{Z}_2^2$ & $\mathbb{Z}^2$ & $0$ \\
\hline
\multirow{2}{*}{$(2,0,0)\sim(0,2,0)\sim(2,2,0)$}
&\multirow{2}{*}{$(C_{s-\delta})^2$}
& AI,D,AII,C  & $\mathbb{Z}^2$ & $0$ & $\mathbb{Z}^2$ & $0$ & $\mathbb{Z}^2$ & $0$ & $\mathbb{Z}^2$ & $0$ \\
&& BDI,DIII,CII,CI & $0$ & $\mathbb{Z}^2$ & $0$ & $\mathbb{Z}^2$ & $0$ & $\mathbb{Z}^2$ & $0$ & $\mathbb{Z}^2$ \\
\hline
\multirow{8}{*}{$(3,0,0)\sim(0,3,0)\sim(3,3,0)$}
&\multirow{8}{*}{$(R_{s+1-\delta})^2$}
& AI  & $\mathbb{Z}_2^2$ & $\mathbb{Z}^2$ & $0$ & $0$ & $0$ & $(2\mathbb{Z})^2$ & $0$ & $\mathbb{Z}_2^2$ \\
&& BDI & $\mathbb{Z}_2^2$ & $\mathbb{Z}_2^2$ & $\mathbb{Z}^2$ & $0$ & $0$ & $0$ & $(2\mathbb{Z})^2$ & $0$ \\
&& D   & $0$ & $\mathbb{Z}_2^2$ & $\mathbb{Z}_2^2$ & $\mathbb{Z}^2$ & $0$ & $0$ & $0$ & $(2\mathbb{Z})^2$ \\
&& DIII& $(2\mathbb{Z})^2$ & $0$ & $\mathbb{Z}_2^2$ & $\mathbb{Z}_2^2$ & $\mathbb{Z}^2$ & $0$ & $0$ & $0$ \\
&& AII & $0$ & $(2\mathbb{Z})^2$ & $0$ & $\mathbb{Z}_2^2$ & $\mathbb{Z}_2^2$ & $\mathbb{Z}^2$ & $0$ & $0$ \\
&& CII & $0$ & $0$ & $(2\mathbb{Z})^2$ & $0$ & $\mathbb{Z}_2^2$ & $\mathbb{Z}_2^2$ & $\mathbb{Z}^2$ & $0$ \\
&& C   & $0$ & $0$ & $0$ & $(2\mathbb{Z})^2$ & $0$ & $\mathbb{Z}_2^2$ & $\mathbb{Z}_2^2$ & $\mathbb{Z}^2$ \\
&& CI  & $\mathbb{Z}^2$ & $0$ & $0$ & $0$ & $(2\mathbb{Z})^2$ & $0$ & $\mathbb{Z}_2^2$ & $\mathbb{Z}_2^2$ \\
\hline
\multirow{2}{*}{$
\begin{array}{c}
(2,1,0)\sim(1,2,0)\sim(3,1,0)\\
\sim(1,3,0)\sim(3,2,0)\sim(2,3,0)\\
\end{array}$}
&\multirow{2}{*}{$C_{s+1-\delta}$}
& AI,D,AII,C & $0$&$\mathbb{Z}$ & $0$ & $\mathbb{Z}$ & $0$ & $\mathbb{Z}$ & $0$ & $\mathbb{Z}$ \\
&& BDI,DIII,CII,CI  & $\mathbb{Z}$ & $0$ & $\mathbb{Z}$ & $0$ & $\mathbb{Z}$ & $0$ & $\mathbb{Z}$ & $0$ \\
\hline
\multirow{8}{*}{
$\begin{array}{cc}
(0,0,1)\sim(2,0,1)\sim(0,2,1),\\
(1,0,1)\sim(0,1,1)\sim(3,0,1)\\
\sim(0,3,1)\sim(3,1,1)\sim(1,3,1)\\
\end{array}$}
&\multirow{8}{*}{$R_{s-\delta}$}
& AI   & $\mathbb{Z}$ & $0$ & $0$ & $0$ & $2\mathbb{Z}$ & $0$ & $\mathbb{Z}_2$ & $\mathbb{Z}_2$ \\
&& BDI  & $\mathbb{Z}_2$ & $\mathbb{Z}$ & $0$ & $0$ & $0$ & $2\mathbb{Z}$ & $0$ & $\mathbb{Z}_2$ \\
&& D    & $\mathbb{Z}_2$ & $\mathbb{Z}_2$ & $\mathbb{Z}$ & $0$ & $0$ & $0$ & $2\mathbb{Z}$ & $0$ \\
&& DIII & $0$ & $\mathbb{Z}_2$ & $\mathbb{Z}_2$ & $\mathbb{Z}$ & $0$ & $0$ & $0$ & $2\mathbb{Z}$ \\
&& AII  & $2\mathbb{Z}$ & $0$ & $\mathbb{Z}_2$ & $\mathbb{Z}_2$ & $\mathbb{Z}$ & $0$ & $0$ & $0$ \\
&& CII  & $0$ & $2\mathbb{Z}$ & $0$ & $\mathbb{Z}_2$ & $\mathbb{Z}_2$ & $\mathbb{Z}$ & $0$ & $0$ \\
&& C    & $0$ & $0$ & $2\mathbb{Z}$ & $0$ & $\mathbb{Z}_2$ & $\mathbb{Z}_2$ & $\mathbb{Z}$ & $0$ \\
&& CI   & $0$ & $0$ & $0$ & $2\mathbb{Z}$ & $0$ & $\mathbb{Z}_2$ & $\mathbb{Z}_2$ & $\mathbb{Z}$ \\
\hline
\multirow{8}{*}{$(1,1,1)\sim(2,1,1)\sim(1,2,1)$}
&\multirow{8}{*}{$R_{s-2-\delta}$}
& AI   & $0$ & $0$ & $2\mathbb{Z}$ & $0$ & $\mathbb{Z}_2$ & $\mathbb{Z}_2$ & $\mathbb{Z}$ & $0$ \\
&& BDI  & $0$ & $0$ & $0$ & $2\mathbb{Z}$ & $0$ & $\mathbb{Z}_2$ & $\mathbb{Z}_2$ & $\mathbb{Z}$ \\
&& D    & $\mathbb{Z}$ & $0$ & $0$ & $0$ & $2\mathbb{Z}$ & $0$ & $\mathbb{Z}_2$ & $\mathbb{Z}_2$ \\
&& DIII & $\mathbb{Z}_2$ & $\mathbb{Z}$ & $0$ & $0$ & $0$ & $2\mathbb{Z}$ & $0$ & $\mathbb{Z}_2$ \\
&& AII  & $\mathbb{Z}_2$ & $\mathbb{Z}_2$ & $\mathbb{Z}$ & $0$ & $0$ & $0$ & $2\mathbb{Z}$ & $0$ \\
&& CII  & $0$ & $\mathbb{Z}_2$ & $\mathbb{Z}_2$ & $\mathbb{Z}$ & $0$ & $0$ & $0$ & $2\mathbb{Z}$ \\
&& C    & $2\mathbb{Z}$ & $0$ & $\mathbb{Z}_2$ & $\mathbb{Z}_2$ & $\mathbb{Z}$ & $0$ & $0$ & $0$ \\
&& CI   & $0$ & $2\mathbb{Z}$ & $0$ & $\mathbb{Z}_2$ & $\mathbb{Z}_2$ & $\mathbb{Z}$ & $0$ & $0$ \\
\hline
\multirow{8}{*}{$(2,2,1)$}
&\multirow{8}{*}{$R_{s+4-\delta}$}
& AI   & $2\mathbb{Z}$ & $0$ & $\mathbb{Z}_2$ & $\mathbb{Z}_2$ & $\mathbb{Z}$ & $0$ & $0$ & $0$ \\
&& BDI  & $0$ & $2\mathbb{Z}$ & $0$ & $\mathbb{Z}_2$ & $\mathbb{Z}_2$ & $\mathbb{Z}$ & $0$ & $0$ \\
&& D    & $0$ & $0$ & $2\mathbb{Z}$ & $0$ & $\mathbb{Z}_2$ & $\mathbb{Z}_2$ & $\mathbb{Z}$ & $0$ \\
&& DIII & $0$ & $0$ & $0$ & $2\mathbb{Z}$ & $0$ & $\mathbb{Z}_2$ & $\mathbb{Z}_2$ & $\mathbb{Z}$ \\
&& AII  & $\mathbb{Z}$ & $0$ & $0$ & $0$ & $2\mathbb{Z}$ & $0$ & $\mathbb{Z}_2$ & $\mathbb{Z}_2$ \\
&& CII  & $\mathbb{Z}_2$ & $\mathbb{Z}$ & $0$ & $0$ & $0$ & $2\mathbb{Z}$ & $0$ & $\mathbb{Z}_2$ \\
&& C    & $\mathbb{Z}_2$ & $\mathbb{Z}_2$ & $\mathbb{Z}$ & $0$ & $0$ & $0$ & $2\mathbb{Z}$ & $0$ \\
&& CI   & $0$ & $\mathbb{Z}_2$ & $\mathbb{Z}_2$ & $\mathbb{Z}$ & $0$ & $0$ & $0$ & $2\mathbb{Z}$ \\
\hline
\multirow{8}{*}{$(3,2,1)\sim(2,3,1)\sim(3,3,1)$}
&\multirow{8}{*}{$R_{s+2-\delta}$}
& AI   & $\mathbb{Z}_2$ & $\mathbb{Z}_2$ & $\mathbb{Z}$ & $0$ & $0$ & $0$ & $2\mathbb{Z}$ & $0$ \\
&& BDI & $0$ & $\mathbb{Z}_2$ & $\mathbb{Z}_2$ & $\mathbb{Z}$ & $0$ & $0$ & $0$ & $2\mathbb{Z}$ \\
&& D    & $2\mathbb{Z}$ & $0$ & $\mathbb{Z}_2$ & $\mathbb{Z}_2$ & $\mathbb{Z}$ & $0$ & $0$ & $0$ \\
&& DIII & $0$ & $2\mathbb{Z}$ & $0$ & $\mathbb{Z}_2$ & $\mathbb{Z}_2$ & $\mathbb{Z}$ & $0$ & $0$ \\
&& AII  & $0$ & $0$ & $2\mathbb{Z}$ & $0$ & $\mathbb{Z}_2$ & $\mathbb{Z}_2$ & $\mathbb{Z}$ & $0$ \\
&& CII  & $0$ & $0$ & $0$ & $2\mathbb{Z}$ & $0$ & $\mathbb{Z}_2$ & $\mathbb{Z}_2$ & $\mathbb{Z}$ \\
&& C    & $\mathbb{Z}$ & $0$ & $0$ & $0$ & $2\mathbb{Z}$ & $0$ & $\mathbb{Z}_2$ & $\mathbb{Z}_2$ \\
&& CI   & $\mathbb{Z}_2$ & $\mathbb{Z}$ & $0$ & $0$ & $0$ & $2\mathbb{Z}$ & $0$ & $\mathbb{Z}_2$ \\
\hline \hline
\end{tabular}
}
\label{tab:table_2U}
\end{center}
\end{table}\restoregeometry

\section{Conclusion}

In this paper, we have systematically studied the classification of topological insulators and superconductors with multiple additional $\mathbb{Z}_2$ point-group symmetries. 
By employing suspension isomorphisms, we showed that higher-dimensional classification groups can be reduced to lower-dimensional data, thereby revealing a hierarchical structure of the classification. 
Furthermore, for both real and complex AZ classes, we derived a general classification formula in the presence of an arbitrary number of unitary $\mathbb{Z}_2$ point-group symmetries, and demonstrated that the classification groups are uniquely determined by a finite set of $\mathbb{Z}_4$ and $\mathbb{Z}_2$ parameters. 
As a concrete example, we carried out detailed computations for the case of $\mathbb{Z}_2^{\times 2}$ point-group symmetry, and confirmed the effectiveness of the method by presenting the complete classification tables. 
These results contribute to a systematic understanding of higher-order topological insulators and superconductors realized under multiple $\mathbb{Z}_2$ point-group symmetries, and serve as a practical framework for explicit classification in the presence of crystalline symmetries.

\section*{Acknowledgements}
We thank Daichi Nakamura for pointing out errors in the first arXiv version of this manuscript, which have been corrected in the present version.
We were supported by JST CREST Grant No. JPMJCR19T2, and JSPS KAKENHI Grant No. JP22H05118 and JP23H01097.

\bibliography{refs}

\end{document}